\documentclass[11pt]{cernrep} \usepackage{graphicx}
\begin{document}
\title{Proposal for a New Reconstruction Technique for SUSY Processes
at the LHC}
\author{M.M. Nojiri$^1$, G. Polesello$^2$ and D.R. Tovey$^3$}
\institute{$^1$ YITP, Kyoto University, Kyoto 606-8502, Japan \\
$^2$ INFN, Sezione di Pavia, Via Bassi 6, 27100 Pavia, Italy \\ $^3$ Department of Physics
and Astronomy, University of Sheffield, Hounsfield Road, Sheffield S3
7RH, UK}
\maketitle
\begin{abstract}
When several sparticle masses are known, the kinematics of SUSY decay
processes observed at the LHC can be solved if the cascade decays
contain sufficient steps. We demonstrate four examples of this full
reconstruction technique applied to channels involving leptons, namely
a) gluino mass determination, b) sbottom mass determination, c) LSP
momentum reconstruction, and d) heavy higgs mass determination.
\end{abstract}

\section{INTRODUCTION}

The potential of the LHC for SUSY parameter determination has been
studied in great detail for the past seven years \cite{atltdr,Hinchliffe:1997iu,Hinchliffe:1999zc,Bachacou:1999zb,
Allanach:2000kt,Abdullin:1998pm}. One of the most
promising methods involves the selection of events from a single decay
chain near the kinematic endpoint. Information on the masses involved
in the cascade decay can be extracted from the endpoint
measurements. It has been established that one can achieve a few
percent accuracy for sparticle mass reconstruction using this
technique with sufficient statistics.

In this paper we propose a new method for reconstructing SUSY events
which does not rely only on events near the endpoint. Instead one
kinematically solves for the neutralino momenta and masses of heavier
sparticles using measured jet and lepton momenta and a few mass
inputs.

To illustrate the idea we take the following cascade decay chain
\begin{equation}
\tilde{g}\rightarrow \tilde{b}b\rightarrow \tilde{\chi}^0_2 bb
\rightarrow \tilde{\ell}bb\ell \rightarrow \tilde{\chi}^0_1bb\ell\ell.
\label{eq1}
\end{equation}
This decay chain is approximately free from SM background with
appropriate cuts. The five SUSY particles which are involved in the
cascade decay have five mass shell conditions;
\begin{eqnarray}
m^2_{\tilde{\chi}^0_1}&=& p^2_{\tilde{\chi}^0_1},\cr
m^2_{\tilde{\ell}}&=& (p_{\tilde{\chi}^0_1}+ p_{\ell_1})^2,\cr
m^2_{\tilde{\chi}^0_2}&=& (p_{\tilde{\chi}^0_1}+
p_{\ell_1}+p_{\ell_2})^2,\cr
m^2_{\tilde{b}}&=& (p_{\tilde{\chi}^0_1}+
p_{\ell_1}+p_{\ell_2}+p_{b_1})^2,\cr
m^2_{\tilde{g}}&=& (p_{\tilde{\chi}^0_1}+ 
p_{\ell_1}+p_{\ell_2}+p_{b_1}+p_{b_2})^2.
\label{gluino}
\end{eqnarray}
Of these five masses, $m_{\tilde{\chi}^0_1}$,$m_{\tilde{\chi}^0_2}$
and $m_{\tilde{\ell}}$ can be measured at the LHC using first
generation squark cascade decays with an accuracy of $\sim$ 10\% (the
mass difference is measured more precisely).  Moreover, with input
from a future high energy Linear Collider these masses might be
determined with an accuracy $\sim O(1\%)$.
We therefore assume for the present work that the masses of the two
lighter neutralinos and of the right handed slepton are known, and we
ignore the corresponding errors.

For a $bb\ell\ell$ event, the equations contain six unknowns
($m_{\tilde{g}}$, $m_{\tilde{b}}$ and $p_{\tilde{\chi}^0_1}$) which
satisfy five equations. For two $bb\ell\ell$ events, we have ten
equations while we only have ten unknowns (two neutralino four
momenta, $m_{\tilde{g}}$ and $m_{\tilde{\chi}^0_1}$ ). Mathematically,
one can obtain the sbottom and gluino masses and all neutralino
momenta if there are more than two $bb\ell\ell$ events.

We call this technique the ``mass relation method'' as one uses the
fact that sparticle masses are common for events which go though the
same cascade decay chain.  Note events need not be near the endpoint
of the decay distribution to be relevant to the mass determination. In
the next section we demonstrate the practical application of this
method to measurement of the masses of the gluino and sbottom.

As a byproduct of the technique, once the mass of the squark and of
all the sparticles involved in the decay are known, the momentum of
the lighter neutralino can be fully reconstructed,
and this further constrains the event.

In SUSY events sparticles are always pair produced and there are two
lightest neutralinos in the event. If squark decays via
$\tilde{q}\rightarrow \tilde{\chi}^0_2\rightarrow
\tilde{\ell}\rightarrow \tilde{\chi}^0_1$ can be identified on one
side of the event then the neutralino momentum can be reconstructed as
described above.  The transverse momentum of the lightest neutralino
in the other cascade decay can then be obtained using the following
equation
\begin{equation}
{\bf p_T}(\tilde{\chi}^0_1(2)) = {\bf p_T}({\rm miss}) +
{\bf p_T}(\tilde{\chi}^0_1(1)), 
\label{ptmiss}
\end{equation}
provided that there are no hard neutrinos involved in the decay.  This
transverse momentum can be used to constrain the cascade decay of the
other sparticle.

For the case where the other squark decays via $\tilde{q}\rightarrow
\tilde{\chi}^+_1q \rightarrow \tilde{\chi}^0_1q W$ followed by
$W\rightarrow q'q''$, the chargino mass can be determined by using
Eq. (\ref{ptmiss}) and the following relations,
\begin{eqnarray}
p_{\tilde{q}}&=& p_{\tilde{\chi}^0_1(2)}+p_j+p_W,
\cr
p^2_{\tilde{q}}&=&m^2_{\tilde{q}},
\end{eqnarray}
where $p_j$ is the momentum of the selected high $p_T$ jet which comes
from the squark decay and $p_W$ is the momentum of the two jet system
consistent with the $W$ interpretation.  The neutralino momentum
resolution is important for the chargino mass reconstruction and we
discuss this in section 3. The reconstruction will be discussed in a
separate contribution\cite{tovey}.

The full reconstruction technique can be extended for higgs mass
reconstruction.  In section 4, we discuss the heavy higgs mass
determination from the process $H\rightarrow
\tilde{\chi}^0_2\tilde{\chi}^0_2$ followed by $\tilde{\chi}^0_2
\rightarrow \tilde{\ell} \ell\rightarrow ll\tilde{\chi}^0_1$. This
process is also useful for discovery of heavy higgs bosons. The four
lepton momenta and missing momentum can be used to reconstruct the
higgs mass assuming that the $p_T$ of the higgs boson is very small.

\section{GLUINO CASCADE DECAY}

We first discuss the results of a simulation study of the process
where a gluino cascade decays into a sbottom at model point
SPS1a\cite{Allanach:2002nj}. The relevant sparticle masses for this
study are listed in Table 1. The events were generated using the
HERWIG 6.4 generator \cite{Corcella:2000bw} \cite{Moretti:2002eu} and
passed through ATLFAST \cite{ATLFAST}, a parametrised simulation of
the ATLAS detector.

\begin{table}
\begin{center}
\begin{tabular}{|c|c|c|c|c|}
\hline
$m_{\tilde{g}}$ & $m_{\tilde{b}_1(2)}$ & $m_{\tilde{\chi}^0_2}$ & 
$m_{\tilde{\ell}_R}$ 
&$m_{\tilde{\chi}^0_1}$ \cr
\hline
595.2& 491.9(524.6)& 176.8&  136.2& 96.0\cr
\hline
\end{tabular}
\caption{Some sparticle masses in GeV at SPS1a. }
\end{center}
\end{table}

We study only events which contain the cascade decay shown in
Eq.(\ref{eq1}). We then apply the following preselections to reduce
backgrounds:
\begin{itemize}
\item $p_T^{miss}>100$~GeV
\item $M_{\rm eff}>600$~GeV
\item at least 3 jets with $p_{T1}>150$~GeV, $p_{T2}>100$~GeV and
$p_{T3}>50$~GeV. 
\item exactly two jets with $p_T>50$~GeV tagged as $b$-jets
\item exactly two OS-SF leptons with $p_{Tl1}>20$~GeV, $p_{Tl2}>10$~GeV, and
invariant mass $40$GeV$<m_{ll}<78$~GeV.
\end{itemize}

The solution of Eq. (\ref{gluino}) can be written in the following
form:
\begin{eqnarray}
m^2_{\tilde{g}}&=& F_0+ F_1 m^2_{\tilde{b}}\pm  F_2 D, \cr
{\rm where} \ \ 
D^2 &\equiv& D_0+ D_1 m^2_{\tilde{b}} +D_2 m^4_{\tilde{b}}.
\label{sol}
\end{eqnarray}
Here $F_i$ and $D_i$ depend upon $p_{\ell_i}$ and $p_{b_i}$ and the
neutralino and slepton masses.  In the event, there are two $b$ jets
and we assume that the $b$ jet with larger $p_T$ originates from the
$\tilde{b}$ decay.  The two leptons must come from $\tilde{\chi}^0_2$
and $\tilde{\ell}$ decay.  There are maximally four sets of gluino and
sbottom mass solutions together with two lepton assignments for each
decay, because we cannot determine from which decay the lepton
originates. To reduce combinatorics we take the event pair which
satisfies the following conditions:
\begin{itemize}
\item  Only one lepton assignment has a solution to the Eq. (\ref{sol}) 
\item For a pair of events there are only two solutions and there is a
difference of more than 100~GeV between the two gluino mass solutions.
\end{itemize}

\begin{figure}[thb]
\begin{center}
\includegraphics[width=5cm]{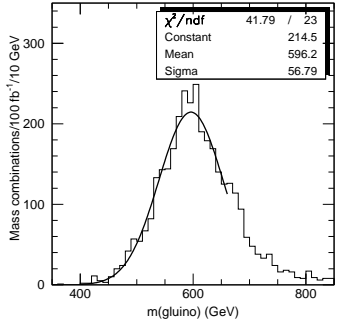}
\end{center}
\caption{$m_{\tilde{g}}$ obtained by using Eq. (\ref{sol})
for two $bb\ell\ell$ events.}
\end{figure}

In Fig 1, we plot the minimum $m_{\tilde{g}}$ solution which satisfies
the conditions given above. The peak position is consistent with the
gluino mass, and the error on the peak position obtained by a Gaussian
fit is around 1.7~GeV for 100 fb$^{-1}$.  For the events used in the
reconstruction, each event is used on average five times.  Note that
the $\sigma$ of the Gaussian fit is large ($\sim 56.7$~GeV) and is
determined by the resolution on the momentum measurement of the four
$b$-jets.  It is worth stressing that the results presented here were
produced by using a parametrised simulation of the response of the
ATLAS detector to jets, based on the results of a detailed simulation.
Results which crucially depend on the detailed features of the
detector response, such as the possibility of discriminating the two
sbottom squarks (see below) need to be validated by an explicit
detailed simulation of the detector performed on the physics channel
of interest. We only attempt here to evaluate the impact of the new
technique on sparticle reconstruction.

Once the gluino mass has been determined one can reconstruct the
sbottom mass by fixing the gluino mass to the measured value.  Here
one needs to solve only Eq.(\ref{sol}), which involves only two $b$-jets
in the fit, and therefore errors due to the jet resolution are
expected to be less than those for the gluino mass reconstruction.

For each event, there are two sbottom mass solutions
$m_{\tilde{b}}$(sol1) and $m_{\tilde{b}}$(sol2), each sensitive to the
gluino mass input.  The difference between the gluino and sbottom mass
solutions is however stable against variation in the assumed gluino
mass. The mass itself may have a large error in the absolute scale,
but the mass differences are obtained rather precisely, as is the case
in the endpoint method.

In Fig. 2 (left), we plot the solutions for all possible lepton
combinations in the $m_{\tilde g}-m_{\tilde{b}}$ (sol1) $m_{\tilde{g}}
-m_{\tilde{b}}$ (sol2) plane. Here we use the $b$-parton momentum
obtained from generator information.  One of the solutions tends to be
consistent with the input sbottom mass. Moreover the two decay modes
$\tilde{g}\rightarrow \tilde{b}_1b$ and $\tilde{b}_2b$ are clearly
separated.

We can compare the results from the previous analysis with those from
the endpoint analysis\cite{LHCLC}, where one uses approximate the
formula
\begin{equation}
{\bf p}_{\tilde{\chi}^0_2}=
\left(1-\frac{m_{\tilde{\chi}^0_1}}{m_{\ell\ell}}\right){\bf p}_{\ell\ell}.
\label{wrong}
\end{equation}
This formula is correct only at the endpoint of the three body decay
$\tilde{\chi}^0_2\rightarrow\chi^0_1 \ell\ell$, but is nevertheless
approximately correct near the edge of $\tilde{\chi}^0_2\rightarrow
\tilde{\ell}\ell \rightarrow \ell\ell\tilde{\chi}^0_1$ for SPS1a.  The
sbottom mass obtained by using Eq.(\ref{wrong}) is shown in
Fig. 2(right). For this case, the $\tilde{b}_2$ peak at 70.6~GeV is
not separated from the $\tilde{b}_1$ peak at 103~GeV.

\begin{figure}[thb]
\begin{center}
\includegraphics[width=6.5cm]{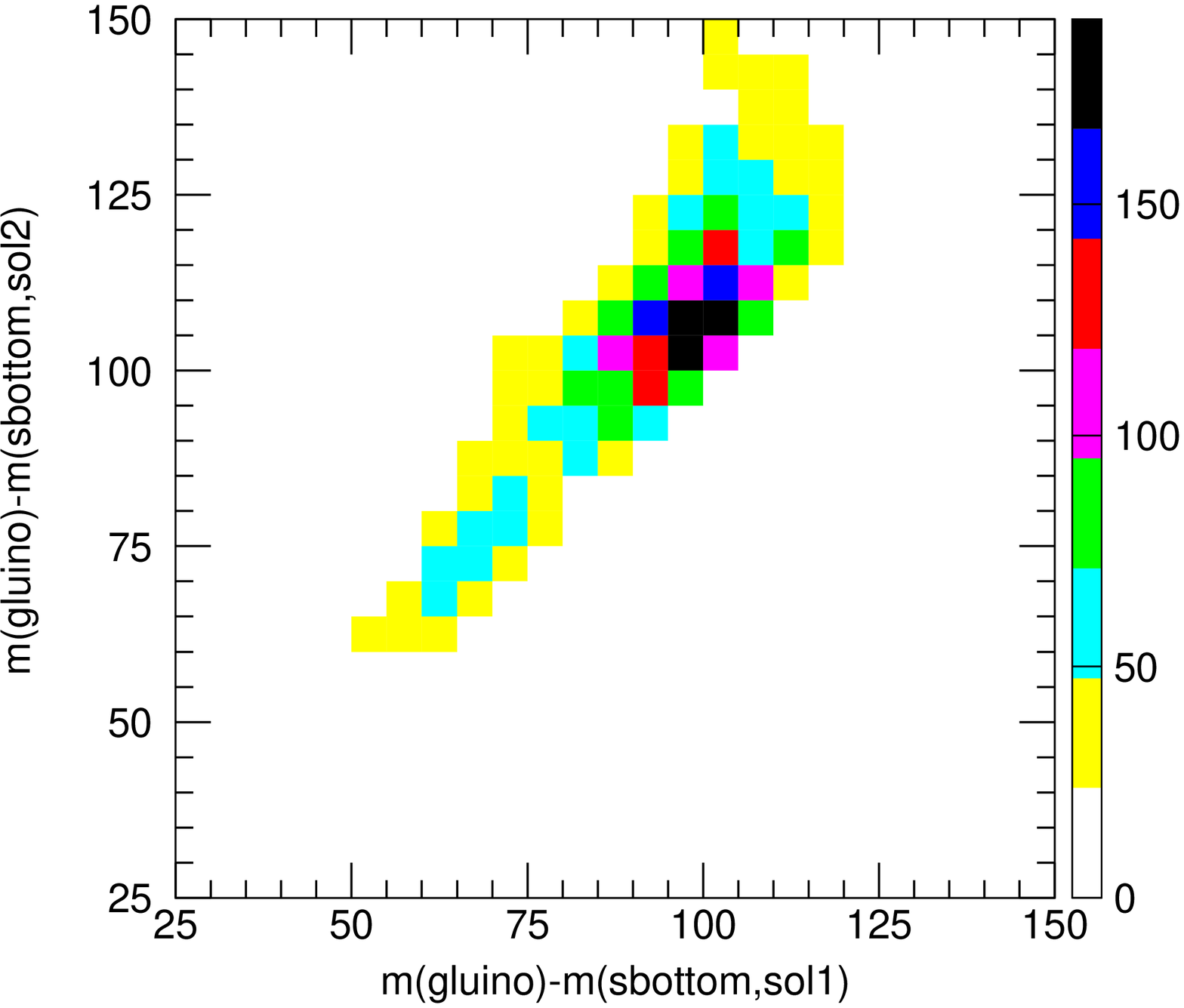}
\includegraphics[width=6.5cm]{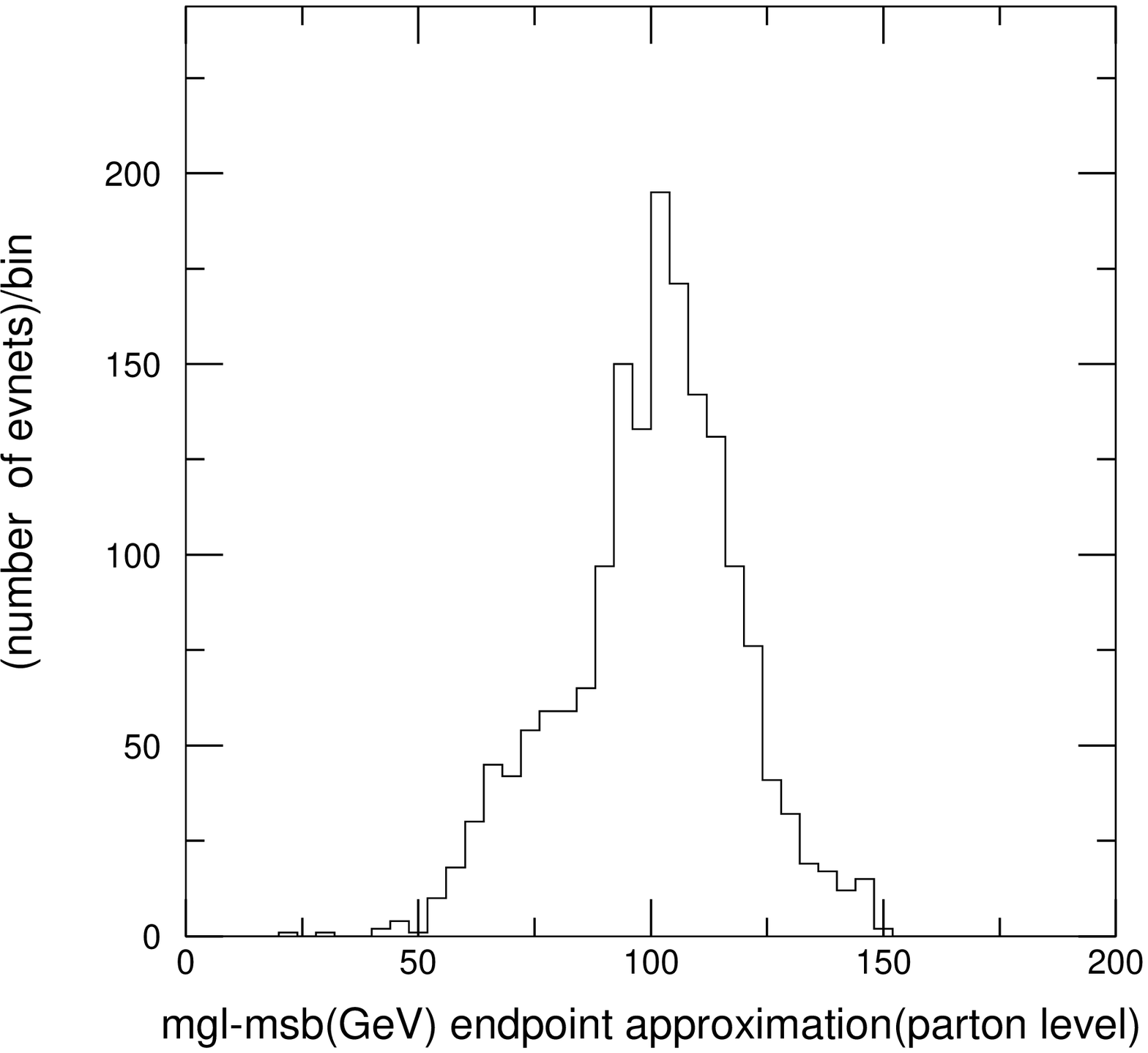}
\end{center}
\caption{The distribution of $m_{\tilde{g}}- m_{\tilde{b}}$ calculated
using the parton level $b$ momentum by solving Eq.(\ref{gluino})
(left) and using the approximate relation Eq. \ref{wrong}(right).}
\end{figure}

\begin{figure}[thb]
\begin{center}
\includegraphics[width=6cm]{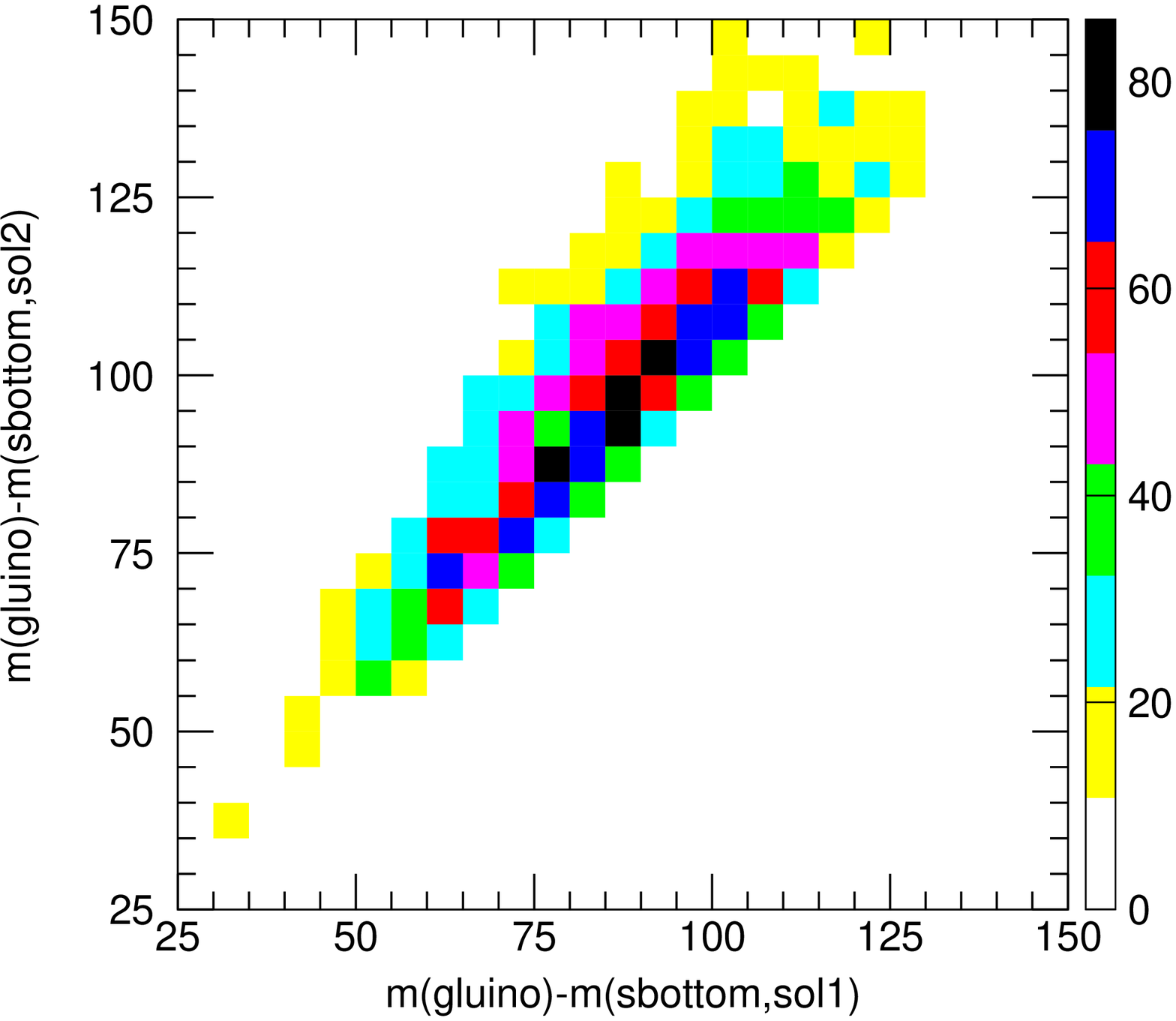}
\hskip 1cm 
\includegraphics[width=6cm]{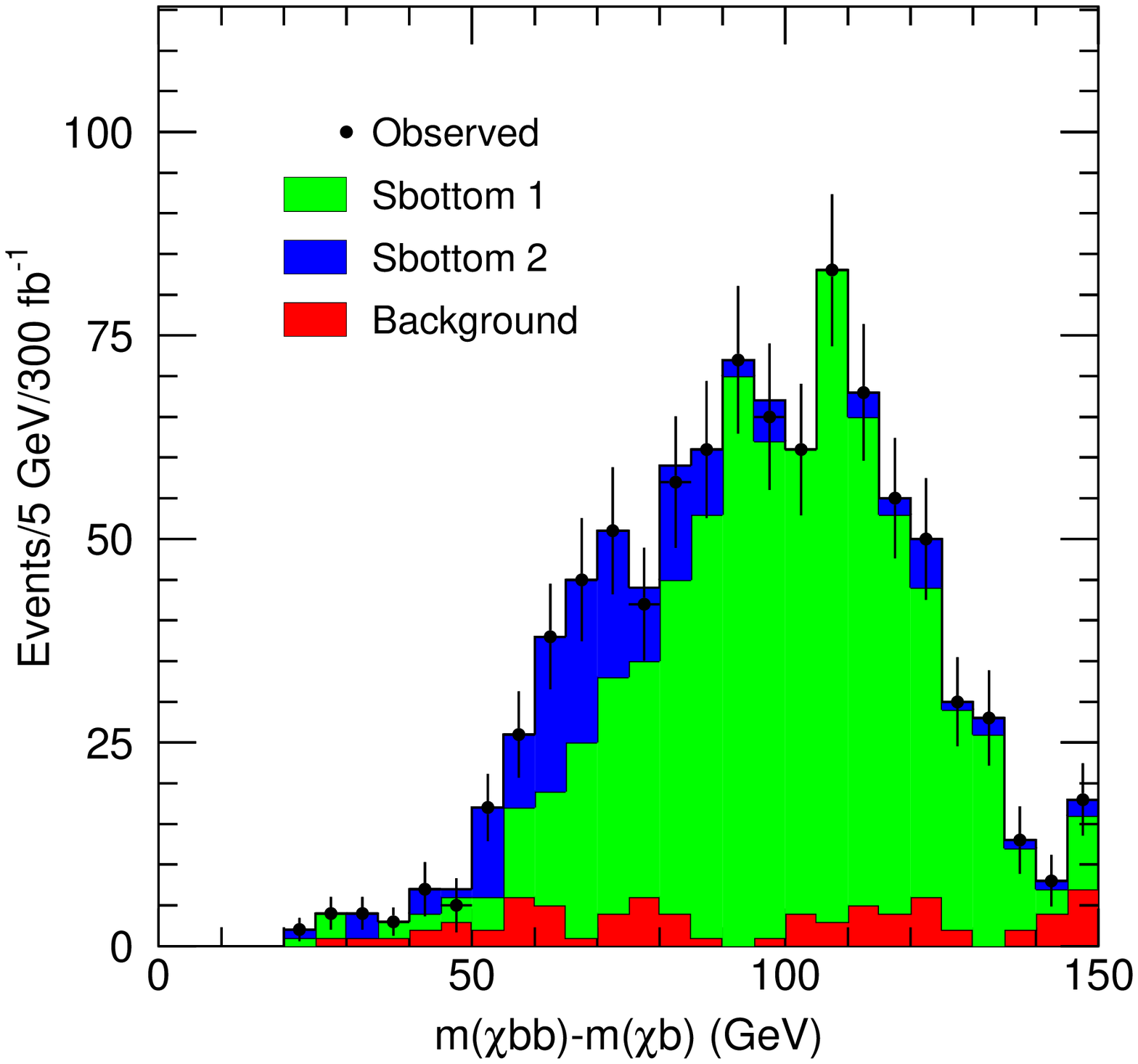}
\end{center}
\caption{As for Fig. 2 but with the $b$ jet momentum used instead of
the $b$ parton momentum.}
\end{figure}

The $\tilde{b}_1$ mass, or the weighted average of the sbottom masses,
is easily obtained.  The $b$ jet resolution is not sufficient however
to clearly separate the $\tilde{b}_1$ and $\tilde{b}_2$.  This can be
seen in Fig. 3 where the plots show the distributions corresponding to
Fig.2(left) and (right) but now with the $b$ parton momenta replaced
by $b$ jet momenta.  For the endpoint analysis (Fig.3 right), 
a correct evaluation of the sbottom masses would require 
a fit taking into account the shape of the response of ATLAS to b-jets.
In order to approximately evaluate the achievable statistical precision,
a naive double gaussian fit was performed on the distribution
shown in Fig.3 right, which corresponds to $\int dt L=300$ fb$^{-1}$.
The resulting statistical uncertainties are 
$\pm 1$~GeV ($ \pm 2.5$~GeV) for the
$m_{\tilde{g}}-m_{\tilde{b}_{1}}$($m_{\tilde{g}}-m_{\tilde{b}_{2}}$ )
peak positions respectively. Additional systematic uncertainties, 
not yet evaluated, as well a 1\% error due to the uncertainty on the
jet energy scale should also be considered.
These numbers are obtained assuming the presence of two gaussian peaks
in the data.

For the mass relation method the number of events available for the
study is larger by a factor of 2 because events away from the
endpoints can be used.  We also use the exact formula for the mass
relation method.  Although the analysis is more complicated due to the
multiple solutions, we believe it to be a worthwhile technique for use
when attempting to reconstruct the $\tilde{b}_1$ and $\tilde{b_2}$
masses.

\section{NEUTRALINO MOMENTUM RECONSTRUCTION} 

In this section, we discuss the reconstruction of the momentum of the
lightest neutralino. As we have discussed already, the mass shell
condition can be solved for long decay cascades, such as
$\tilde{q}\rightarrow \tilde{\chi}^0_2 q\rightarrow
\tilde{\ell}q\ell\rightarrow \tilde{\chi}^0_1q\ell\ell$.  For this
process we have two neutralino momentum solutions for each lepton
assignment.  One may wonder if the solutions for the neutralino
momentum might be smeared significantly, because of the worse jet energy
resolution as compared to leptons, and the jet $p_T$
is generally much larger than the neutralino momentum for the cascade
decay.  In Fig.  4(left) we show the distribution of
$p_T$(reco)/$p_T$(truth) for the point studied in \cite{tovey}.  Here
we choose the correct lepton combination using generator information,
and take the solution which minimizes $\vert
p_T$(reco)/$p_T$(truth)$-1\vert$.  Except for the case where we took
the wrong jet as input the reconstructed $p_T$ is within 20\% of the
true neutralino momentum. The result for the gluino cascade decay into
sbottom Eq.(1) is similar.
 
In Fig. 4(right) we show a similar reconstruction for the gluino
cascade decay, but unlike Fig.4(left), we use both lepton
combinations.  We fix the gluino mass to the input value\footnote {Here
we adopt an event selection which makes use of the true (input) gluino
and sbottom mass values, although in practice fitted values would be
used.}  and take events where one of the four sbottom mass solutions
is consistent with the input sbottom mass such that $\vert
m_{\tilde{b}_1}-m_{\tilde{b}}({\rm best})\vert<10$~GeV.  We then take
the solution where the sbottom mass is closest
to the input $m_{\tilde{b}_1}$.  There are still two
$p_{\tilde{\chi}^0_1}$ solutions, and we choose the one which 
minimize $\vert \min(p_{T}({\rm reco})/p_T({\rm truth}), 
p_{T}({\rm truth})/p_T({\rm reco}))
-1\vert$.   
The neutralino momentum resolution is
worse than that obtained using the correct lepton assignments only.
Nevertheless a significant fraction of events are reconstructed with
$0.8<\vert {\bf p}({\rm reco})/ {\bf p}({\rm truth})\vert <1.2$.

\begin{figure}[thb]
\begin{center}
\includegraphics[width=4.5cm]{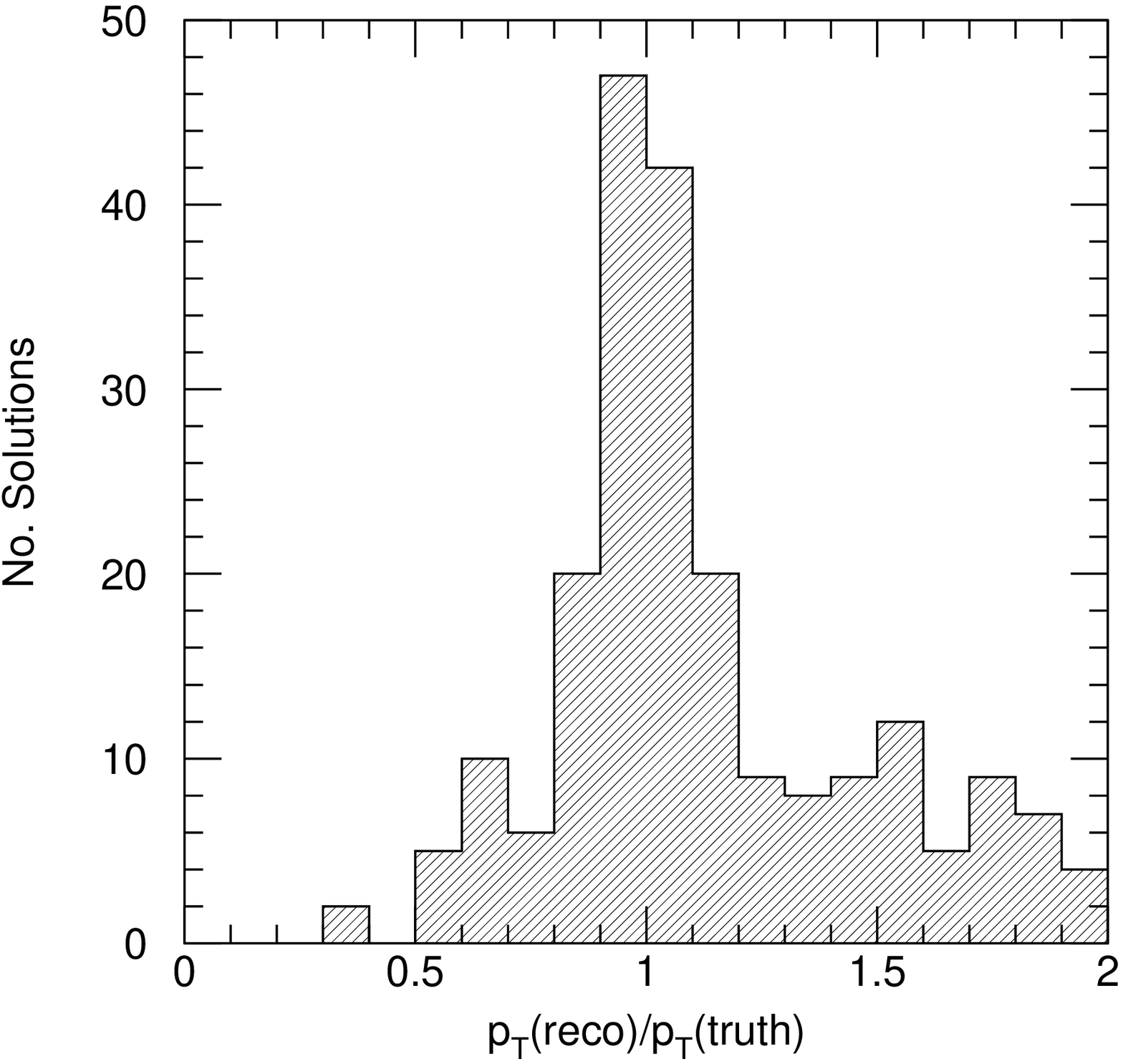}
\includegraphics[width=4.5cm]{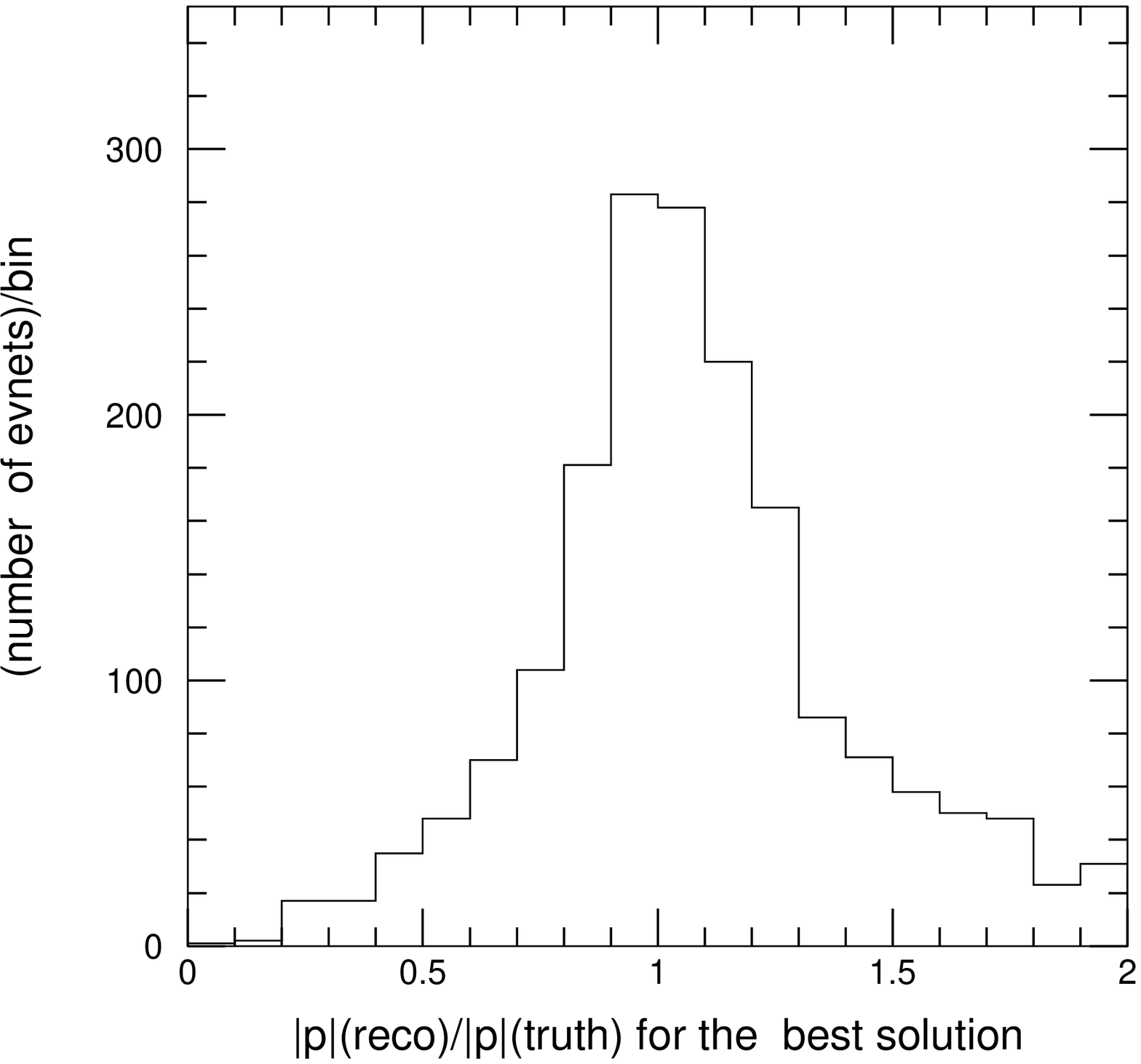}
\end{center}
\caption{Left: The calculated $\tilde{\chi}^0_1$ transverse momentum
divided by the true transverse momentum.  The decay
$\tilde{q}_L\rightarrow \tilde{\chi}^0_2 q \rightarrow \tilde{\ell}
q\ell \rightarrow\tilde{\chi}^0_1q \ell\ell$ is studied for the model
point used for the chargino study: $m_0=100$~GeV, $m_{1/2}=300$~GeV,
$A_0=-300$~GeV, $\tan\beta=6$, and $\mu>0$.  Only the correct lepton
choice is used.  Right: $\vert {\bf p}$(reco)$/{\bf p}$(truth)$\vert$
for the decay chain Eq.(\ref{eq1}) for SPS1a.}
\end{figure}

\section{HIGGS MASS RECONSTRUCTION}
\def\tchi{\tilde{\chi}} \def\tl{\tilde{\ell}} \def\etmiss{E_T^{miss}}
A promising decay for the observation of heavy and pseudo-scalar higgs
bosons in the difficult region with intermediate $\tan\beta$ is the
decay into two neutralinos. When both neutralinos decay through the
chain
$$
\tchi^0_2\rightarrow\tl_R\ell\rightarrow\ell\ell\tchi^0_1
$$
the resulting signature consists of events with four isolated leptons
(paired in opposite-sign same-flavour pairs) and no jet activity. The
main SM backgrounds to this signature are $t\bar{t}$ production, where
both the $b$-jets and the $W$s decay into leptons and $Zbb$
production. The key element for the rejection of these backgrounds is
the fact that the leptons from $b$ decays are not isolated. A detailed
study of the performance of lepton isolation in the detector is needed
to assess the visibility of the signal.  Additionally there is an
important SUSY background, including irreducible backgrounds from
direct slepton and gaugino decay.  Full background
studies as a function of the SUSY parameters were  performed by the
ATLAS and CMS Collaborations \cite{TDR,filip}.  
We propose here, along the lines of
the previous sections, a technique for the complete reconstruction of
the higgs peak, based on the knowledge of the masses of $\tchi_2^0$,
$\tl_R$ and $\tchi_1^0$.  In this case one has 8 unknown quantities:
the 4-momenta of the two LSP's, and 8 constraints: six on-shell mass
constraints (3 for each leg), and the two $\etmiss$ components.

To demonstrate the power of the method, we apply it to Point SPS1a,
for which the mass of the $A$ and of the $H$ is $\sim394$~GeV.  The BR
into $\tchi^0_2\tchi^0_2$ is 6\% (1\%) for the $A$($H$).  We perform
the study on 1000 events for
$$A\rightarrow\tchi^0_2\tchi^0_2\rightarrow
\ell\ell\ell\ell\tchi^0_1\tchi^0_1$$ corresponding approximately to
the expected statistics for 300 fb$^{-1}$.  We simply require 2
isolated leptons with $p_T>20$~GeV and 2 further isolated leptons
with $p_T>10$~GeV, all within $|\eta|<2.5$.  The efficiency of these
cuts is $\sim 60\%$.

\begin{figure}[thb]
\begin{center}
\includegraphics[width=6cm]{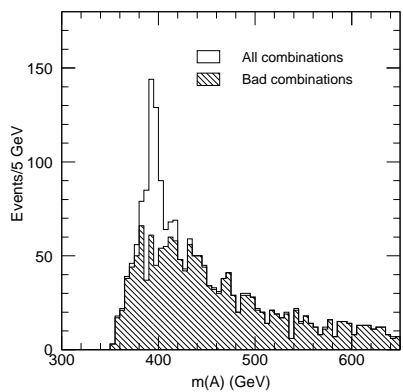}
\end{center}
\caption{Reconstructed mass of the $A$.} 
\label{fig:higgs}
\end{figure}

We have not performed any background simulations because at this stage
we only wish to explore the viability of the full reconstruction
technique.  The main problem for the reconstruction is the correct
assignment of the leptons to the appropriate decay chain.  The first
selection is based on requiring a unique identification of the lepton
pairs coming from the decays of the two $\tchi^0_2$s.  We therefore
require that either of the following two criteria is satisfied:
\begin{itemize}
\item
the flavour configuration of the leptons is $e^+e^+\mu^+\mu^-$
\item
the lepton configuration is either $e^+e^-e^+e^-$ or
$\mu^+\mu^-\mu^+\mu^-$, but for one of the two possible pairings the
invariant mass of one of the pairs is larger than 78~GeV, i.e.  above
the lepton-lepton edge for the $\tchi^0_2$ decay.
\end{itemize}
The total efficiency after these cuts is $\sim30\%$.  At this point,
on each of the two legs there is still an ambiguity due to the fact
that each lepton can be either the product of the first or of the
second step in the decay chain. This gives 4 possible combinations.
Furthermore, the full reconstruction results in a quartic equation
which can have zero, two or four solutions.  We show in
Fig.~(\ref{fig:higgs}) the distribution of the calculated $A$ mass for
all of the retained combinations as a full line. The dashed line shows
the combinations with the wrong lepton assignment.  A clear and narrow
peak emerges over the combinatorial background.  The width is
approximately 6~GeV, determined by the resolution of the measurement
of the momentum of the leptons.

\section{CONCLUSIONS}
In this contribution, we have described a novel technique for
reconstructing the mass and momenta of SUSY particles.  This technique
does not rely on any approximate formulae nor on endpoint
measurements. All events contribute to the sparticle mass
determination and decay kinematics reconstruction, even if they are
away from the endpoint of the distribution.  The method may be
particularly useful when the SUSY mass scale is large. In that case
the statistics can be so low that the endpoint cannot be seen clearly
while the SUSY sample itself is very clean.

The method applies most effectively when we know some of the
sparticles' masses exactly, because the number of unknown parameters
in e.g.  Eq.(\ref{gluino}) is reduced.  In the particular case where
some of the sparticle masses are measured at a LC the sparticle
cascades may be solved completely and study of the decay distributions
and higher mass determination becomes possible at LHC.

When all the sparticle masses are known the neutralino momentum can be
reconstructed if four sparticles are involved in the cascade
decay. The sparticles would be pair produced, and if we can identify
both of the cascade decay chains in the events then we only need six
sparticles in the cascade decay to solve both of the neutralino
momenta on account of the missing momentum constraint. The
reconstruction of sparticle momenta provides us with an interesting
possibility for studying the decay distribution at the LHC.

On the other hand, our method is not valid when some of the particles
in the cascade produce hard neutrinos.  This is unfortunately the case
when the chargino decays into (s)leptons, when a $\tilde{\tau}$ is
involved in the decay, or when a $W$ is produced and decays
leptonically.  If such SUSY decay processes dominate then this method
may not be useful.

\section*{Acknowledgments}
This work was performed in the framework of the workshop: Les Houches
2003: Physics at TeV Scale Colliders. We wish to thank the staff and
organisers for all their hard work before, during and after the
workshop.  We thank members of the ATLAS Collaboration for helpful
discussions. We have made use of ATLAS physics analysis and simulation
tools which are the result of collaboration-wide efforts. DRT wishes
to acknowledge PPARC and the University of Sheffield for support.

\bibliography{new6}

\end{document}